\documentclass[runningheads]{svmult}
\usepackage{graphicx}
\begin{document}
\title*{Extremal Properties of Random Structures}
\toctitle{Extremal Properties of Random Structures}
\titlerunning{Extremal Properties of Random Structures}
\author{E.~Ben-Naim\inst{1}\and P.~L.~Krapivsky\inst{2} \and
S.~Redner\inst{2}}
\authorrunning{Ben-Naim, Krapivsky, Redner}
\institute{Theoretical Division and Center for Nonlinear
Studies, Los Alamos National Laboratory, Los Alamos, NM 87545 \and
Center for Polymer Studies and Department of Physics, Boston
University, Boston, MA 02215}
\maketitle

\begin{abstract}
  The extremal characteristics of random structures, including trees,
  graphs, and networks, are discussed.  A statistical physics approach
  is employed in which extremal properties are obtained through
  suitably defined rate equations.  A variety of unusual time
  dependences and system-size dependences for basic extremal
  properties are obtained.
\end{abstract}

\section{Introduction}

The goal of this article is to show that methods of non-equilibrium
statistical physics are very useful for analyzing extreme properties
of random structures.  Extremes are compelling human curiosities ---
we are naturally drawn to compilations of various pinnacles of
endeavor, such as lists of the most beautiful people, the richest
people, the most-cited scientists, athletic records, {\it etc}
\cite{guiness}. More importantly, extremes often manifest themselves
in catastrophes, such as the failure of space shuttles, the breaching
of dams in flood conditions, or stock market crashes.  The theory of
extreme statistics \cite{ejg,jg,res} is a powerful tool for describing
the extremes of a set of independent random variables; however, much less 
is known about extremes of correlated variables \cite{cl,dm,rcps}.  Such an
understanding is crucial, since complex systems are composed of many
subsystems that are highly correlated.

While estimates for the failure probability of a nuclear plant or a
space shuttle still involve guesswork, understanding the extremes of
certain correlated random variables is a hard science.  Below we
demonstrate this thesis for various extremal characteristics of
geometrical features in basic evolving structures, such as randomly
growing trees, graphs, and networks.  In each case, the growth process
of the structure induces correlations in the variables whose extremes
are the focus of this review.  We shall illustrate how the statistical
physics of classical irreversible processes can be naturally adapted
to elucidate both typical and extremal statistics.

We obtain new scaling laws for extreme properties and consequently
give new insights for a variety of applications.  For example, random
trees arise naturally in data storage algorithms \cite{hmm,dek,ws}, an
important branch of computer science, and the maximal branch height
yields the worst-case performance of data retrieval algorithms.
Random trees also describe various non-equilibrium processes, such as
irreversible aggregation \cite{mvs,sc} and collisions in gases
\cite{vvd}.  Random graphs \cite{bb,jlr} have numerous applications to
computer science and to physical processes such as polymerization
\cite{pjf}.  Random growing networks are used to model the
distributions of biological genera, word frequencies, and income
\cite{yule,simon}, the structure of the Internet \cite{willinger}, the
World-Wide Web \cite{applics}, and social networks \cite{sp,gn}.

As a subtext to this review, it is worth mentioning that problems at
the interface of statistical physics and computer science have been
fruitful and symbiotic.  Algorithms and methods developed in one area
have found application in the other field; important examples include
the Monte Carlo method, simulated annealing, and the Dijkstra
algorithm.  Statistical physics concepts such as criticality, scaling,
universality, and techniques such as replicas have proved useful in
diverse interdisciplinary applications such as algorithmic complexity,
combinatorial optimization, error correction, compression algorithms, and
image restoration; a review of these topics can be found in
Refs.~\cite{mmz,hn,mpz,jb,fps,ks}.

We will focus on three ubiquitous random structures --- trees, graphs,
and networks.  Random trees (Sec.~2) can be viewed as the space-time
diagram of irreversible aggregation with a size-independent merging
rate.  This connection allows to apply well-known results in
aggregation to elucidate the growth of the largest component (the
leader) and the number of changes in its identity. The number of lead
changes grows quadratically with logarithm of the system size.  The
time-dependent number of lead changes becomes asymptotically
self-similar, following a scaling form in which the scaling variable
involves a logarithmic, rather than an algebraic ratio, between the
typical size and the system size.  Qualitatively similar properties
also characterize the smallest component in the system.

Another characteristic of random trees is their height.  The
corresponding branch height distribution is Poissonian, reflecting the
random nature of the merger process that underlies tree growth.  The
growth of the tree height (the maximal branch height) has an
interesting relation to traveling wave propagation.  The velocity of
this wave yields typical and extremal height statistics as a
corollary.

Random graphs (Sec.~3) are also equivalent to an aggregation process
in which the merging rate of two components is proportional to the
product of their sizes.  This system undergoes a gelation transition
in which a giant component, that contains a finite fraction of the
entire mass in the system, arises.  Near this transition, the size
distribution of graph components follows a self-similar behavior.
Despite the differences with the size distribution of random trees,
leadership statistics in these two systems are remarkably robust.

Random networks (Sec.~4) can be grown by adding nodes and attaching
the new node to a pre-existing node with a rate that depends on the
degree of the target node.  A hallmark of such systems is the
statement that ``the rich get richer''; that is, the more popular
nodes tend to remain so.  This adage is in keeping with human
experience --- a person who is rich is likely to stay rich, as
evidenced by the continued appearance of the same individuals on lists
of wealthiest individuals \cite{wealth}.  We examine whether the rich
really get richer in growing networks by studying properties of the
nodes with the largest degree.  In keeping with analysis of random
trees and graphs, we focus on the identity of the most popular node as
a function of time, the expected degree of this most popular node, and
the number of lead changes in the most popular node as a function of
time.

\section{Random Trees}

Random trees underlie physical processes such as coagulation,
collisions in gases \cite{vvd}, and the fragmentation of solids
\cite{ho,km}.  They are also important in computer science algorithms
such as data storage and retrieval \cite{hmm,dek,bp,ld,pn,mk1}.
Different extremal characteristics may be important in different
contexts.  In aggregation processes, the maximal aggregate size is of
interest.  In other cases, the maximal or the minimal branch height
are of interest. In Lorentz gases, the maximum branch height is
related to the largest Lyapunov exponent, while in data storage,
extremal heights yield best-case and worst-case algorithm
performances.

\begin{figure}[ht]
\centerline{\includegraphics*[width=5cm]{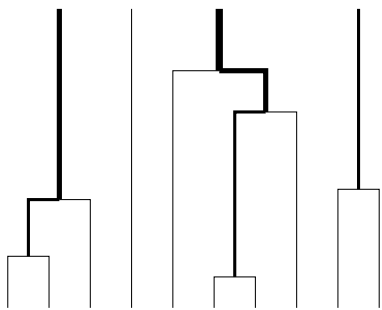}}
\caption{Random trees.  A forest of random growing trees is equivalent
  to the space-time evolution of irreversible aggregation with a
  size-independent merging rate.  Each branch corresponds to the world
  line of a cluster.  The thickness of each branch is proportional to
  the size of the cluster.  The sizes of the 4 trees are (left to
  right) 3, 1, 4, 2 and their heights are 2, 0, 3, 1.}
\label{tree}
\end{figure}

Consider a forest of random trees that is generated randomly as
follows (Fig.~\ref{tree}).  Starting with $N$ single-branch trees, two
trees are picked randomly and merged.  This process is repeated until
a single tree containing all $N$ branches is generated.  We treat the
merger process dynamically.  Let $k$ be the number of trees. The
transition $k\to k-1$ occurs with rate $r_k$ proportional to the total
number of pairs. Choosing $2/N$ as the merger rate for each pair
(i.e., $r_k=k(k-1)/N$) is convenient as in the thermodynamic limit
$N\to\infty$, the normalized density $c=\langle k\rangle/N$ evolves
according to $\frac{d}{dt}c=-c^2$. Given the initial condition
$c(0)=1$, the density is
\begin{equation}
\label{ct-rt}
c(t)=\frac{1}{1+t}.
\end{equation}
The number of trees is therefore ${\cal N}=\langle k\rangle
=N(1+t)^{-1}$. Moreover, conservation of the total number of branches
yields the average tree size $m=1+t$. The results are stated in terms
of the physical time $t$, but can be easily re-expressed in terms of
the intrinsic quantities ${\cal N}$ or $m$.

\subsection{Size Statistics}

Let $n_k(t)$ be the number of trees with $k$ branches at time $t$. The
normalized density $c_k(t)=n_k(t)/N$ evolves according to the Smoluchowski
rate equation \cite{mvs,sc,da}
\begin{equation}
\label{ckt-eq-rt}
\frac{dc_k}{dt}=\sum_{i+j=k}c_ic_j-2cc_k
\end{equation}
with the monodisperse initial conditions $c_k(0)=\delta_{k,1}$. This
evolution equation reflects the fact that trees merge randomly,
independent of their size.  The well-known solution to this equation
is
\begin{equation}
\label{ckt-rt}
c_k(t)=\frac{t^{k-1}}{(1+t)^{k+1}}.
\end{equation}
Taking the long time limit $t\to\infty$ while keeping the variable
$k/t$ fixed, the size distribution approaches the asymptotic form
$c_k(t)\to t^{-2}e^{-k/t}$.  More generally, this can be recast as the
scaling form
\begin{equation}
\label{ckt-scl-rt}
c_k(t)\simeq k_*^{-2}\Phi(k/k_*),
\end{equation}
with the scaling function $\Phi(z)=e^{-z}$ and the typical tree size
$k_*\simeq t$.

\subsection{The Leader}

Extremal characteristics, such as the size of the largest tree --- the
leader --- and the number of lead changes, follow directly from the
size distribution.  We focus on the asymptotic time
regime\footnote{The behavior in the early time regime, $t\ll 1$, can
be obtained by using the exact time dependence (\ref{ckt-rt}).}, where
most of the lead changes occur, and use the scaled size distribution
(\ref{ckt-scl-rt}).  Let $l(t,N)$ be the average size of the leader at
time $t$.  The basic criterion used to determine the size of the
leader is
\begin{equation}
\label{basic}
U_l(t)\equiv\sum_{j\geq l} n_j\simeq Nt^{-1}e^{-l/t}=1.
\end{equation}
This simply states that there is one cluster whose size exceeds $l(t,N)$.
Solving for the leader size gives
\begin{equation}
\label{lt-rt}
l(t,N)\simeq t\,\ln\frac{N}{t}.
\end{equation}
This expression holds in the asymptotic time regime $t\gg 1$.  For
short times the leader size grows logarithmically with system size
$l(t\approx 1)\sim \ln N$.  Finally, at times of the order $N$, the
leader becomes of the order of the system size.  The final leader,
that is, the ultimate winner, emerges on a time scale of the order
$N$.  This is consistent with the fact that the average ``final'' time
for a single tree to remain in the system $t_f$, is given by $t_f=N-1$
as follows from ${\cal N}=1$.

We now consider the quantity $L(t,N)$, defined as the average number
of lead changes during the time interval $(0,t)$.  Lead changes occur
when two trees (neither of which is the leader) merge and overtake the
leader.  The flux of probability to surpass the leader is simply the
rate of change of the cumulative distribution.  Thus
$\frac{d}{dt}L(t,N)=\frac{\partial}{\partial t}U_k\big|_{k=l}$.  Using
$U_l=1$ yields $\frac{d}{dt}L(t,N)\simeq lt^{-2}\simeq t^{-1}\ln
\frac{N}{t}$.  Therefore, the time-dependent number of lead changes is
\cite{bk}
\begin{equation}
\label{ltn-rt}
L(t,N)\simeq\ln t\ln N-\frac{1}{2}(\ln t)^2,
\end{equation}
which can be recast in the self-similar form
\begin{equation}
\label{ltn-scl-rt}
L(t,N)\simeq(\ln N)^2\,F(x), \qquad x=\frac{\ln t}{\ln N},
\end{equation}
with the quadratic scaling function $F(x)=x-\frac{1}{2}x^2$.  Notice the
unusual scaling variable --- a ratio of logarithms --- in contrast to the
ordinary scaling variable $z=k/k_*$ underlying the size distribution
(\ref{ckt-scl-rt}).  The scaling variable still involves the typical size,
$x=\ln k_*/\ln N$.  Note also that the leader size (\ref{lt-rt}) can be
expressed in terms of the same scaling variable $l(t,N)\simeq t\ln N\,f(x)$
with $f(x)=\frac{d}{dx}F(x)=1-x$.  Numerical simulations confirm this scaling
behavior \cite{bk}. However, the convergence to these asymptotics is slow due
to the logarithmic functional dependences on the system size and time.

\begin{figure}[t]
\centerline{\includegraphics*[width=6.5cm]{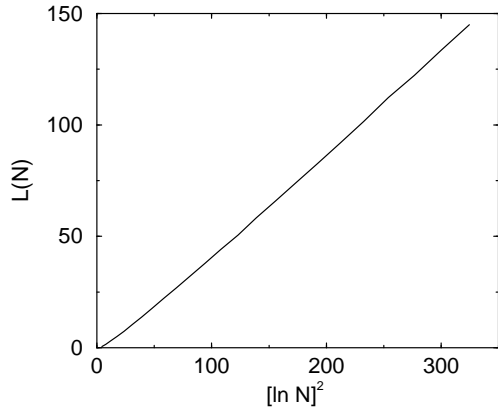}}
\caption{The total number of lead changes $L(N)$ versus the system
size $N$. The simulation data represents an average over $10^4$
independent realizations of the random tree generation process with
$N$ up to $10^8$.}
\label{ln}
\end{figure}

The total number of lead changes $L(N)$ as a function of system size
$N$ follows from the time dependent behavior (\ref{ltn-rt}).  The
eventual winner emerges at time of order $N$. Using $L(N)\cong
L(t\propto N,N)$ we obtain
\begin{equation}
\label{ln-rt}
L(N)\simeq A(\ln N)^2\,
\end{equation}
with $A=F(1)=1/2$ (see Fig.~\ref{ln}). The correction to this leading
asymptotic behavior is of the order $\ln N$.  The logarithmic
dependence implies that lead changes are relatively infrequent.

Both the size of the leader and the number of lead changes grow
logarithmically in the early time regime, $l(t\approx 1,N)\propto
L(t\approx 1,N)\propto \ln N$. The first relation implies that
initially the leader size predominantly grows in increments of one and
every leader is a new leader.  When $t\gg 1$, the size of the leader
greatly exceeds the number of lead changes as the increments of the
leader size grow roughly linearly with time.

The distribution of the number of lead changes $P_n(t,N)$, {\it i.e.},
the probability that $n$ lead changes occur by time $t$, can be
determined by noting that lead changes occur by a random process in
which the average flux of probability to surpass the leader is
$\frac{d}{dt}L$.  Hence, the probability distribution obeys
\hbox{$\frac{d}{dt}P_n=(\frac{d}{dt}L)\,[P_{n-1}-P_n]$} with the
initial condition $P_n(0,N)=\delta_{n,0}$.  Therefore, the
distribution of the number of lead changes is Poissonian and it is
characterized solely by the average number of lead changes
\begin{equation}
\label{pn-rt}
P_n(t,N)=\frac{[L(t,N)]^n}{n!}\,e^{-L(t,N)}.
\end{equation}
As a result, the ultimate number of lead changes is also Poissonian
distributed, \hbox{$P_n(N)=\frac{L^n}{n!}\,e^{-L}$}, with $L\equiv
L(N)$ given by Eq.~(\ref{ln-rt}).  Asymptotically, the Poissonian
distribution approaches a Gaussian in the proximity of the peak:
\begin{equation}
\label{pn-gauss}
P_n(N)\simeq \frac{1}{\sqrt{2\pi L}}\,\exp\left[-\frac{(n-L)^2}{2L}\right].
\end{equation}
The number of lead changes is a self-averaging quantity; however, the
system size should be huge to ensure that relative fluctuations
$\frac{\delta n}{n}\sim \frac{\sqrt{L}}{L}\sim (\ln N)^{-1}$ are
small.  Hence in a given realization for a system of size $N=10^8$
(the maximum size in our simulations), lead changes are still
relatively erratic.

\begin{figure}[t]
\centerline{\includegraphics*[width=6.5cm]{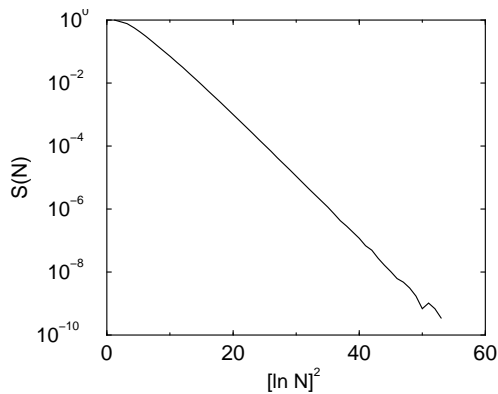}}
\caption{The survival probability of the first leader $S(N)$ versus
the system size $N$ obtained from an average over $10^{10}$
independent realizations. The slope $A=1/2$ is in accordance with
Eq.~(\ref{sn-rt}).}
\label{sn}
\end{figure}

Another interesting quantity is $S(N)$, the probability that no lead
change ever occurs.  This is obviously the ``survival'' probability
that the first leader, whose size is initially $k=2$, never
relinquishes the lead.  This survival probability is given by
$S(N)\equiv P_0(N)=\exp(-L)$, so it decays faster than a power-law but
slower than a stretched exponential (Fig.~\ref{sn})
\begin{equation}
\label{sn-rt}
S(N)\simeq \exp\left[-A(\ln N)^2\right].
\end{equation}

The above formalism extends to the statistics of the $r^{\rm
th}$-largest tree.  Using $U_l=r$, the average size of the $r^{\rm
th}$-largest tree grows according to $l_r\simeq t\ln
\frac{N}{rt}$. Moreover, the total number of changes in the group of
$r$-largest trees grows linearly with $r$ according to
$\frac{r}{2}(\ln N)^2$.

Among several open problems we mention just two: What is the size of
the winner (the last emerging leader)?  At what time does the winner
emerge? The averages of both these random quantities grow linearly
with $N$, but we do not know the proportionality factors.  The
computation of these factors, and the determination of the
distribution of these random quantities, are interesting open
problems.

\subsection{The Laggard}

At the opposite end of the size spectrum sits the laggard, the
smallest component in the system.  Unlike the leader, the laggard does
not change its size for a relatively long period.  From the expression
for the monomer density $n_1=N(1+t)^{-2}$, we see that monomers are
depleted from the system only when the time becomes of the order of
$N^{1/2}$. Until this time, the laggard size remains unity.  To
investigate laggard statistics in the interesting regime $N^{1/2}\ll
t\ll N$ we employ the same approach as for the leader. First, we
estimate the cumulative distribution $u_k=\sum_{j=1}^k n_j$ and find
$u_k\simeq t^{-1}(1-e^{-k/t})$.  Then we use the criterion $u_\ell=1$
and get the average laggard size
\begin{equation}
\label{lt-lag}
\ell(t,N)\simeq-t\ln\left(1-\frac{t}{N}\right).
\end{equation}
In the time regime $N^{1/2}\ll t\ll N$, the above expression
simplifies to $\ell(t,N)\simeq t^2/N$.  As in the leader case, the
laggard size is proportional to the typical size, but modified by a
logarithmic correction.

The number of changes in the identity of the laggard, ${\cal L}(t,N)$,
is given by $\frac{d}{dt}{\cal L}(t,N)=-\frac{\partial}{\partial
t}u_k\big|_{k=l}$. Using asymptotics for $u$ and $\ell$, we simplify
the right-hand side and obtain $\frac{d}{dt}{\cal
L}=t^{-1}-Nt^{-2}(1-t/N)\ln (1-t/N)\simeq 2t^{-1}$ for $t\ll N$.
Integrating over time and recalling that the first laggard change
occurs at time of the order $N^{1/2}$ we obtain ${\cal L}(t,N)\simeq
\int_{N^{1/2}}^t ds\,\, 2s^{-1}$.  Consequently,
\begin{equation}
\label{ltn-lag}
{\cal L}(t,N)\simeq 2\ln t -\ln N.
\end{equation}
This behavior can be recast in the scaling form ${\cal L}(t,N)\simeq
(\ln N)F(x)$ with the same scaling variable as in the leader problem,
$x=\ln t/\ln N$, and the linear scaling function $F(x)=2x-1$.  The
total number of laggard changes saturates at
\begin{equation}
\label{ln-lag}
{\cal L}(N)\simeq \ln N.
\end{equation}
Numerical simulations confirm this behavior.  Thus, the total number
of laggard changes is much smaller compared with the leader.  This
behavior is intuitive: it is more difficult to catch up with the rest
of the pack than it is to remain ahead of the pack.

The distribution of the number of laggard changes is also Poissonian,
as in (\ref{pn-rt}).  Moreover, the survival probability still decays
exponentially with the total number of changes ${\cal
S}(N)=\exp[-{\cal L}(N)]$.  However, the growth of the average is only
logarithmic in this case, so the survival probability decays as a
power law
\begin{equation}
\label{sn-lag}
{\cal S}(N)\sim N^{-1},
\end{equation}
{\it i.e.}, much slower than in the leader case.  This can be
understood by considering the probability that the laggard remains a
monomer until the very last merger event between the final two
subtrees.  Interestingly, the size distribution of these final two
trees is uniform as can be seen immediately by considering the
time-reversed merger process.  The probability that the laggard in the
last merging event is a monomer is simply $2/(N-1)$.  This lower bound
for the survival probability is indeed consistent with (\ref{sn-lag}).
An interesting open question is the size distribution of the loser
(the final laggard).

\subsection{Height Statistics}

The height (or depth) of a tree branch provides another fundamental
size characterization.  It is defined as the number of different-width
line segments between a branch and the tree root (see
Fig.~\ref{tree}).  Thus, different heights correspond to different
branches in the tree.  It is therefore natural to ask: What is the
typical branch height?  What is the typical tree height (the maximal
branch height)? What is the maximal tree height?

First, consider the distribution of branch heights.  Each time two
branches merge, the distance to the root increases by one (the branch
height can also be viewed as the generation number).  Let $h(t)$ be
the average number of merger events experienced by a given branch up
to time $t$.  The rate of growth of the average height is proportional
to the number density of trees and since the merger rate equals $2$,
we have $\frac{d}{dt}h(t)=2c(t)$, with $h(0)=0$.  Therefore, the
average branch height is $h(t)=2\ln (1+t)$, or, in terms of the
average tree size $m$,
\begin{equation}
\label{ht} h=2\ln m.
\end{equation}
Thus, the branch height grows logarithmically with its size.  Because
the merger process is random, the probability $P_n(t)$ that the branch
height equals $n$ is Poissonian
\begin{equation}
\label{ph}
P_n(t)={[h(t)]^n\over n!}e^{-h(t)},
\end{equation}
with $h(t)$ the average height.

The height of a tree is defined as the maximal branch height.  For
example, the (left-to-right) trees in Fig.~\ref{tree} have heights of
$2$, $0$, $3$, and $1$, respectively.  Based on the branch height
behavior, we anticipate that the tree height grows logarithmically,
$H_{\rm max}\simeq v_{\rm max}\ln m$.  Similar to the calculation of
the maximal size from the cumulative distribution, the tree height can
be obtained heuristically from the properly normalized branch height
distribution $c^{-1}P_n$ via $\sum_{n\geq H_{\rm max}}
c^{-1}P_n=1$. Estimating the tails of the Poisson distribution
(\ref{ph}) by using the Stirling formula leads to the transcendental
equation \cite{bkm}
\begin{equation}
\label{veq}
v\ln {2e\over v}=1.
\end{equation}
The larger root of this equation yields the growth of the tree height
\begin{equation}
\label{hmax} H_{\rm max}\simeq v_{\rm max}\ln m,\qquad v_{\rm
max}\cong4.31107.
\end{equation}
This value was obtained in different contexts, including fragmentation
processes \cite{ho,km}, and collision processes in gases, where this
value is related to the largest Lyapunov exponent \cite{vvd}.

Each tree carries a height $k$.  The result of a merger between trees
with heights $i$ and $j$ is a tree with height ${\rm max}(i,j)+1$.
The number density of trees with height $k$, $H_k(t)$, evolves
according to the master equation (the initial conditions are
$H_k(0)=\delta_{k,0}$)
\begin{equation}
\label{ck-max}
\frac{dH_k}{dt}=H_{k-1}^2-2c\,H_k+ \sum_{j=k}^\infty
H_j+2H_{k-1}\sum_{j=0}^{k-2}H_j.
\end{equation}

\begin{figure}[ht]
\centerline{\includegraphics*[width=6.5cm]{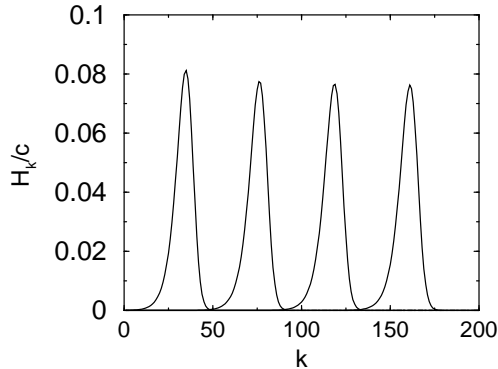}}
\caption{The traveling wave. Shown is the normalized distribution
$c^{-1}H_k$ vs. $k$ at different times $\tau=10,20,30,40$ obtained
from numerical integration of the rate equation (\ref{cumulative}).}
\label{wave-max}
\end{figure}

The rate equations (\ref{ck-max}) are more complicated than the {\it
recursive} Smoluchowski equations (\ref{ckt-eq-rt}) for the tree size
distribution.  Fortunately, one can extract analytically almost all
relevant information without explicitly solving
Eqs.~(\ref{ck-max}). Fig.~\ref{wave-max} shows that the normalized
distribution $c^{-1}H_k$ approaches a traveling wave in the large time
limit. This suggests seeking an asymptotic solution of the traveling
wave; this construction therefore greatly simplifies our analysis. The
traveling wave form has significant qualitative implications for the
tree height statistics, {\it e.g.}, fluctuations with respect to the
mean saturate to some fixed value.

The equations simplify using the cumulative fractions
$A_k=c^{-1}\sum_{j=0}^k H_j$ and the time variable $\tau=\int_0^t
dt'\,c(t')=\ln(1+t)$.  With these transformations, Eqs.~(\ref{ck-max})
become
\begin{equation}
\label{cumulative}
\frac{dA_k}{d\tau}=A_{k-1}^2-A_k
\end{equation}
with the initial conditions $A_k(0)=\delta_{k,0}$.  Substituting the
traveling wave solution, $A_k(\tau)\to A(k-v\tau)$, into
(\ref{cumulative}) we find that $A(x)$ satisfies the nonlinear
difference-differential equation
\begin{equation}
\label{wave}
vA'(x)=A(x)-A^2(x-1)
\end{equation}
with the boundary conditions $A(-\infty)=0$ and $A(\infty)=1$.  This
nonlinear and nonlocal equation appears insoluble; however, important
physical features can now be established analytically.  For example,
both extreme tails of $A(x)$ are exponential:
\begin{equation}
\label{tails} A(x)\sim \cases{e^{x/v}&$x\to-\infty$;\cr
1-e^{-\lambda x}&$x\to\infty$.}
\end{equation}
Consequently, the distribution of both very large and very small
(compared with the typical) heights are exponential. The propagation
velocity of the wave, which characterizes the typical behavior,
follows from the large-$k$ tail.  Substituting $1-A(x)\sim e^{-\lambda
x}$ into (\ref{wave}) gives a dispersion relation, {\it i.e.}, a
relation between the velocity $v$ and the decay constant $\lambda$:
\begin{equation}
\label{dispersion}
v=\frac{2e^{\lambda}-1}{\lambda}.
\end{equation}
Out of the spectrum of possible $v$ only one value, the maximal
possible velocity, is selected\footnote{This actually happens for a
wide class of initial conditions including all that vanish for
sufficiently small $k$.}.  {}From (\ref{dispersion}) we find $v_{\rm
max}\cong 4.31107$, corresponding to $\lambda\cong 0.768039$.  This
velocity satisfies (\ref{veq}) and is identical to the one obtained
heuristically (\ref{hmax}).  Numerical integration shows that a
traveling wave is indeed approached (Fig.~\ref{wave-max}) and the
predicted propagation velocity is confirmed to within $0.1\%$.  The
choice of the extremal velocity is the fundamental selection principle
that applies to classical reaction-diffusion equations
\cite{jdm,mb,wvs,bd,evs} and to numerous difference-differential
equations \cite{mk}.

The traveling wave form of the height distribution implies that the
height --- the elemental random variable --- is highly concentrated
near the average; more precisely, each moment $\langle (H_k-\langle
H_k\rangle)^n\rangle$ is finite. Thus, accurate determination of the
average is especially important.  We already know that $\langle
H_k\rangle\cong v\tau$; a more sophisticated traveling wave
technique yields the leading (logarithmic) correction: $\langle
H_k\rangle\cong v\tau-\frac{3}{2\lambda}\,\ln\tau$ \cite{bkm}.

Similar analysis can also be performed for the minimal branch height
\cite{bkm}.  The resulting velocity $v_{\rm min}\cong 0.373365$ is the
smaller root of the transcendental equation (\ref{veq}).

\subsection{The Tallest and the Shortest}

The tallest tree is defined as the one with largest height and
similarly for the shortest tree. The tallest and the shortest are
merely the height leader and laggard, respectively.  The number of
changes in the identity of these extremal trees throughout the
evolution process follows from the tails of the height distribution.

\begin{figure}[ht]
\centerline{\includegraphics*[width=6.5cm]{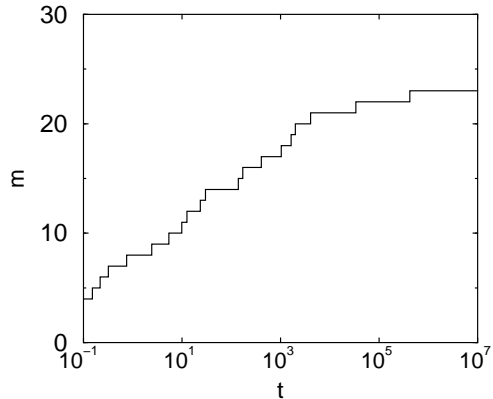}}
\caption{Number of times $m$ the tallest tree changes versus time $t$
in a single realization with $N=10^7$.}
\label{tallest}
\end{figure}

Consider the height distribution $H_k(t)$ and the corresponding
cumulative distribution $U_k=N\sum_{j\geq k} H_j(t)$.  Both of these
distributions have exponential tails\footnote{The proportionality
factor is tacitly ignored as it is irrelevant asymptotically.  The
determination of its value requires a nonlinear analysis of the
traveling wave.}, $U_k(t)\sim NH_k(t)\sim Nt^{\lambda
v-1}\exp(-\lambda k)$, as follows from the large-$x$ tail of the
traveling wave (\ref{tails}).  The criterion $U_l=1$ yields the
average height of the tallest tree
\begin{equation}
l(t,N)\simeq \lambda^{-1}\ln [Nt^{\lambda v-1}].
\end{equation}
Indeed, the height of the tallest tree saturates at a time scale of
the order $N$ consistent with the saturation value $l_{\rm
final}(N)\simeq v\ln N$. This is also an upper bound for the total
number of lead changes since the height of the tallest tree grows by
increments of unity. Similar to the leader, $L(t\approx 1)\sim
l(t\approx 1) \sim \ln N$. However, at later times the rate of change
is slower, $\frac{d}{dt}L(t)\sim t^{-1}$, as follows from the flux
criterion $\frac{d}{dt}L=\frac{\partial}{\partial
t}U_k\big|_{k=l}$. The overall number of changes now grows slower than
in the leader case $L(N)\simeq\varphi\ln N$ with $\varphi\leq v_{\rm
max}$ and consequently, the survival probability of the first tallest
tree decays algebraically
\begin{equation}
S(N)\sim N^{-\varphi}
\end{equation}
with an apparently non-trivial exponent $\varphi$.  Determination this
constant is challenging since the number of lead changes in the early
and the late time regimes are comparable. Nevertheless, this heuristic
approach successfully yields extremal statistics of an extremal tree
characteristic, namely, the maximal branch height. The irregular
nature of the lead changing process is manifest when a single
realization is considered (Fig.~\ref{tallest}).

Extremal statistics of the shortest tree follow from the cumulative
distribution $u_k=N\sum_{j\leq k}H_j\sim Nt^{-2}\exp(k/v)$ and the
criteria $u_{\ell}=1$ and $\frac{d}{dt}{\cal L} =-\frac{\partial
}{\partial t}u_k\big|_{k=\ell}$.  The size of the smallest tree thus
grows according to
\begin{equation}
\ell(t,N)\simeq v\ln \frac{t^2}{N}
\end{equation}
for times $t\gg N^{1/2}$ (at earlier times the shortest tree is a
monomer). Even though the shortest tree has a different growth law
than the laggard (\ref{lt-lag}), the time dependent number of changes
grows according to (\ref{ltn-lag}).  Thus the total number of changes
${\cal L}(N)\simeq \ln N$ and the survival probability ${\cal
S}(N)\sim N^{-1}$ are as in the laggard case.

We conclude that leadership statistics generally exhibit logarithmic
dependences on the system size. However, they are not
universal. Different behaviors may characterize leaders and laggards
and the behavior may depend on the type of geometric feature, {\it
i.e.}, size or height.  We have observed both linear and quadratic
growth with $\ln N$.  A third possibility, saturation at a finite
value, is found for random networks, as will be shown below.

\section{Random Graphs}

Random graphs are fundamental in theoretical computer science
\cite{dek,bb,ws,jlr}. They have been used to model social networks
\cite{sp,gn}, and physical processes such as percolation \cite{ds} and
polymerization \cite{pjf}.  We discuss size statistics only.  The size
distribution is derived and then used to obtain leader statistics.

\begin{figure}[ht]
\centerline{\includegraphics*[width=5cm]{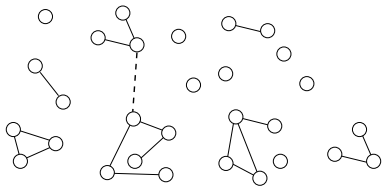}}
\caption{A random graph.  The dashed line indicates a newly-added link
 that joins two randomly-selected nodes.  The probability of joining
 together two components that contain $i$ and $j$ nodes is
 proportional to $ij$.}
\label{graph}
\end{figure}

\subsection{The Size Distribution}

A random graph is grown from an initially disconnected graph with $N$
nodes.  Two nodes are then selected at random and are connected.  This
process occurs at a constant rate, that we set equal to unity without
loss of generality.  This linking is repeated indefinitely until all
$N$ nodes form a single connected component.

Let $n_k$ be the number of components of size $k$.  The normalized
density $c_k=n_k/N$ evolves according to the Smoluchowski equation
\begin{equation}
\label{ckt-eq-rg}
\frac{dc_k}{dt}=\frac{1}{2}\sum_{i+j=k}ijc_ic_j-k\,c_k.
\end{equation}
The initial conditions are $c_k(0)=\delta_{k,1}$.  In writing
(\ref{ckt-eq-rg}), the conservation law $\sum_k kc_k=1$ is employed.
Equations (\ref{ckt-eq-rg}) reflect that components are linked with a
rate proportional to the product of their sizes.

The generating function $F(z,t)=\sum_k kc_k(t)e^{kz}$, evolves
according to ${\partial F\over \partial t}=(F-1) {\partial F\over
\partial z}$ with the initial condition $F(z,0)=e^z$.  Writing the
derivatives through Jacobians, ${\partial F\over \partial t}=
{\partial (F, z)\over \partial (t, z)}$ and ${\partial F\over \partial
z}= {\partial (F, t)\over \partial (z, t)}$, and using the relation
${\partial z\over \partial t}= {\partial (z, F)\over \partial (t,
F)}$, the nonlinear equation for $F(z,t)$ is recast into the linear
equation ${\partial z\over \partial t}=1-F$, from which we
get\footnote{The integration constant $\ln F$ follows from the initial
condition $F(z,0)=e^{z}$.}  $z(t)=(1-F)t+\ln F$.  Exponentiating this
equality gives an implicit relation for the generating functions
\begin{equation}
\label{fzt}
F(z,t)\,e^{-tF(z,t)}=e^{z-t}.
\end{equation}
The Lagrange inversion formula\footnote{The series $v=\sum_{n\ge 1}
  {n^{n-1}\over n!}u^n$ is a solution of the equation $ve^{-v}=u$
  \cite{wilf}.}  conveniently yields the size distribution
  \cite{jbm,hez}
\begin{equation}
\label{ckt-rg}
c_k(t)=\frac{(kt)^{k-1}}{k\cdot k!}e^{-kt}.
\end{equation}
The system undergoes a gelation transition at time $t_g=1$.  At this
point a giant component arises that eventually engulfs the entire mass
in the system.  Close to the gelation time, the size distribution
attains the scaling behavior
\begin{equation}
\label{ckt-scl-rg}
c_k(t)\simeq k_*^{-5/2}\Phi(k/k_*),
\end{equation}
with the scaling function $\Phi(z)=\frac{1}{\sqrt{2\pi}}
z^{-5/2}e^{-z/2}$. The typical size diverges, \hbox{$k_*\simeq
(1-t)^{-2}$}, as $t\to t_g$. Beyond the gelation point, there exists
an infinite sequence of transitions at times $t_k\simeq k^{-1} \ln N$
beyond which components of size $k$ disappear.  At the last such
transition time $t_1$, the system consists of the giant component and
a few surviving monomers.  The smallest component is always a monomer
and the laggard problem is trivial.

\subsection{The Leader}

The size of the giant component (the last emerging leader) follows
from the size distribution. Exactly at the gelation time, the
large-size tail of the size distribution is algebraic,
\hbox{$c_k(t=1)\sim k^{-5/2}$}, so that the cumulative distribution is
\hbox{$u_k\sim Nk^{-3/2}$}.  The criterion $u_{l_w}\sim 1$ gives the
average size of the giant component $l_w\sim N^{2/3}$ \cite{bb} and
the time at which it emerges is $1-t_w\sim N^{-1/3}$.

Consider the size of the leader, $l(t,N)$, and the number of lead
changes $L(t,N)$.  At early times ($t\ll 1$), the behavior is the same
as for random trees: the size of the leader $l(t,N)$, the number of
lead changes $L(t,N)$, as well as the number of distinct leaders are
all of the order $\ln N$.  The asymptotic time regime in this case is
$t\to 1$, as suggested by the size distribution.  The tail of the size
distribution together with $u_l=1$ yield an implicit relation for the
size of the leader, \hbox{$l\simeq 2(1-t)^{-2}\,\ln N-3(1-t)^{-2}\,\ln
l\,$}.  Substituting the zeroth order approximation
\hbox{$l^{(0)}=2(1-t)^{-2}\,\ln N$} into $\ln l$ and ignoring
subdominant $\ln\ln N$ terms gives the leader size
\begin{equation}
\label{lt-rg}
l(t,N)\simeq\frac{2}{(1-t)^2}\,\ln [N(1-t)^3].
\end{equation}
At early stages ($t\ll 1$) the leader size grows logarithmically with
the system size.  Moreover, the leader size is proportional to the
typical size but with a logarithmic enhancement.

The rate of leadership change is estimated as in the random tree case
and we find $\frac{d}{dt}\,L(t,N)=(1-t)\,l(t,N)$, so that the time
dependence of the number of lead changes is
\begin{equation}
\label{rg-lnt}
L(t,N)\simeq 2\,\ln N\,\ln\frac{1}{1-t}
-3\left[\ln \frac{1}{1-t}\right]^2\,.
\end{equation}
It follows that the scaling form is
\begin{equation}
\label{ltn-scl-rg} L(t,N)\simeq (\ln N)^2\,F(x) \qquad
x=\frac{\ln\frac{1}{1-t}}{\ln N},
\end{equation}
with the scaling function $F(x)=2x-3x^2$. This scaling function is
related to the leader size: $l(t,N)\simeq k_*\ln N f(x)$, with
$f(x)=\frac{d}{dx}F(x)=2-6x$. The scaling behavior is obeyed until the
giant component emerges, {\it i.e.}, up to a time $t_w$, with
$1-t_w\sim N^{-1/3}$.  We neglected extremely slowly growing terms
that are of the order $\ln\ln N/\ln N$ to obtain the scaling behavior.
Thus, the approach to the scaling behavior may be very slow.

The total number of lead changes, $L(N)\simeq \frac{1}{3}(\ln N)^2$,
is similar to the random tree case\footnote{The relation $1-t_w\sim
N^{-1/3}$ shows that $x\leq 1/3$, and the prefactor is obtained from
the scaling function: $A=F(1/3)=1/3$.}. Furthermore, the distribution
of lead changes is Poissonian, as in (\ref{pn-rt}), and the survival
probability decays according to (\ref{sn-rt}).

Random trees and random graphs show very different size
characteristics.  Gelation occurs in one case but not in the other.
Nevertheless, leadership statistics in these two systems are
remarkably similar.  In both cases, the total number of lead changes
grows as $L(N)\sim (\ln N)^2$.  Moreover, the seemingly different
scaling variables underlying (\ref{ltn-scl-rt}) and (\ref{ltn-scl-rg})
can be both related to the typical size $x=\ln k_*/\ln N$.

\section{Random Networks}

In the case of sequentially growing networks, the basic quantity is
the degree distribution $N_k$, defined as the number of nodes of
degree $k$ when the network contains $N$ total nodes.  In this
section, we investigate extremal properties of the degree
distribution.  We are again interested in the leader, namely, the node
with the highest degree and its associated statistical properties.

\begin{figure}[ht]
\centerline{\includegraphics*[width=6cm]{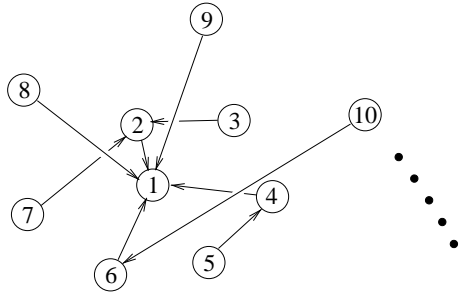}}
\caption{A random network. The network is grown by adding links
  sequentially.  A new node has a single outgoing link that joins to
  an earlier node of degree $k$ with an attachment rate $A_k$.  Each
  node is labeled by its index $J$.}
\label{network}
\end{figure}

\subsection{Identity of the Leader}

We characterize the $J^{\rm th}$ node that enters the network as
having an index $J$ (Fig.~\ref{network}).  To start with an
unambiguous leader node, we initialize the system to have $N=3$ nodes,
with the initial leader having degree 2 (and index 1) and the other
two nodes having degree 1.  A new leader arises when its degree
exceeds that of the current leader.

For a constant attachment rate ($A_k=1$), the average index of the
leader grows algebraically, $J_{\rm lead}(N)\sim N^{\psi}$, with
$\psi\approx 0.41$.  The leader is typically an early node (since
$\psi<1$), but not necessarily one of the very earliest.  For example,
a node with index greater than 100 has a probability of approximately
$10^{-2}$ of being the leader in a graph of $N=10^5$ nodes.  Thus the
order of node creation plays a significant but not deterministic role
in the identity of the leader node for constant attachment rate ---
there is partial egalitarianism.

We can understand this behavior analytically from the joint
index-degree distribution.  Let $C_k(J,N)$ be the average number of
nodes of index $J$ and degree $k$.  For constant attachment rate, this
joint distribution obeys the rate equation
\begin{equation}
\label{ck}
{\partial C_k\over \partial N}
={\partial C_k\over \partial J} +{C_{k-1}- C_k\over N}
+\delta_{k1}\delta(N-J).
\end{equation}
This is a slight generalization of the rate equation for the degree
distribution itself \cite{KR01}.  The new feature is the first term on
the right that accounts for node ``aging''.

The homogeneous form of this equation implies a self-similar solution.
Thus, we seek a solution as a function of the single variable $J/N$
rather than two separate variables \cite{KR01}
\begin{equation}
\label{ck1scal}
C_k(J,N)=f_k(x) \qquad {\rm with}\quad x=\frac{J}{N}.
\end{equation}
This turns Eq.~(\ref{ck}) into the ordinary differential equation
\begin{equation}
\label{fk1}
-x\,{df_k\over dx}=f_{k-1}- f_k.
\end{equation}
We have omitted the delta function term, since it merely provides the
boundary condition $c_k(J=N,N)=\delta_{k,1}$, or
$f_k(1)=\delta_{k,1}$.  The solution is simply the Poisson
distribution in the variable $\ln x$, {\it i.e.},
\begin{equation}
\label{ck0}
C_k(J,N)=\frac{J}{N}\,\frac{|\ln(J/N)|^{k-1}}{(k-1)!},
\end{equation}
{}from which the average index of a node of degree $k$ is
\begin{equation}
\label{av-J}
J_k(N) = \frac{\sum_{1\leq J\leq N} J \,C_k(J,N)}{{\sum_{1\leq J\leq
      N} C_k(J,N)}} = N\left({2\over3}\right)^k.
\end{equation}
Thus the index of the leader is $J_{\rm lead}(N)=N (2/3)^{k_{\rm
max}}$.  The maximum degree is estimated from the extreme value
criterion $\sum_{k\geq k_{\rm max}} N_k(N)\approx 1$ and using
$N_k(N)= N/2^k$ \cite{KR01} gives $k_{\rm max}\sim{\ln N/\ln 2}$.
Therefore \cite{kr}
\begin{eqnarray*}
J_{\rm lead}(N)\sim N^\psi,\qquad {\rm with}\qquad \psi=2-\frac{\ln
3}{\ln 2}\cong 0.415\,037,
\end{eqnarray*}
in excellent agreement with numerical results.

For the linear attachment rate, $A_k=k$, numerical simulations
indicate that a rich gets richer phenomenon arises, as the average
index of the leader $J_{\rm lead}(N)$ saturates to a finite value of
approximately 3.4 as $N\to\infty$.  With probability $\approx 0.9$,
the leader is among the 10 earliest nodes, while the probability
$\approx 0.99$ the leader is among the 30 earliest nodes \cite{kr}.
In general, we find similar behavior for the more general case of the
shifted linear attachment rate $A_k=k+\lambda$.

We can understand these results analytically through the joint
index-degree distribution.  For the linear attachment rate one has
\cite{KR01}
\begin{equation}
\label{ck1all}
C_k(J,N)=\sqrt{J\over N}\left(1-\sqrt{J\over N}\right)^{k-1},
\end{equation}
from which $J_k(N) = 12N/[(k+3)(k+4)]$.  Since $N_k(N) \simeq 4N/k^3$
for linear attachment \cite{BA,KR01}, the extreme statistics criterion
now gives $k_{\rm max}\sim N^{1/2}$.  Therefore $J_{\rm lead}(N)\simeq
12N/ k_{\rm max}^2={\cal O}(1)$ indeed saturates to a finite value.
Thus the leader is one of the first few nodes in the network.

\subsection{Number of Lead Changes}

In contrast with random trees and random graphs, the average number of
lead changes $L(N)$ grows only logarithmically in $N$ for both the
attachment rates $A_k=1$ and $A_k=k$.  While the average number of
lead changes appears to be universal, there is a significant
difference in the distribution of the number of lead changes,
$P_n(N)$, at fixed $N$.  For $A_k=1$, this distribution is sharply
localized, while for $A_k=k$, $P_n(N)$ has a significant large-$n$
tail.  This tail stems from repeated lead changes among the two
leading nodes.  Related to lead changes is the number of {\em
distinct\/} nodes that enjoy the lead over the history of the network.
Simulations indicate that this quantity also grows logarithmically in
$N$.

This logarithmic behavior can be easily understood for the attachment
rate $A_k=1$.  Here the number of lead changes cannot exceed an upper
bound given by the maximal degree $k_{\rm max}\sim \ln N/\ln 2$.  To
establish the logarithmic growth for the general attachment rate
$A_k=k+\lambda$, we first note that when a new node is added, the lead
changes if the leadership is currently shared between two (or more)
nodes and the new node attaches to a co-leader.  The number of
co-leader nodes (with degree $k=k_{\rm max}$) is $N/k_{\rm
max}^{3+\lambda}$, while the probability of attaching to a co-leader
is $k_{\rm max}/N$.  Thus the average number of lead changes satisfies
\begin{equation}
\label{LN}
{dL(N)\over dN}\sim
{k_{\rm max}\over N}\,{N\over k_{\rm max}^{3+\lambda}}.
\end{equation}
Since $k_{\rm max}\sim N^{1/(2+\lambda)}$, Eq.~(\ref{LN}) reduces to ${d
  L/dN}\sim N^{-1}$ and thus gives the logarithmic growth $L(N)\sim \ln N$.

\subsection{Fate of The First Leader}

We now turn to the probability that the first leader retains the lead
throughout the network growth.  For the linear attachment rate
$A_k=k+\lambda$ (rich get richer systems), the initial leader has a
finite chance to remain in the lead forever.  However, for the
egalitarian attachment rate $A_k=1$, the initial leader is eventually
replaced by another leader.  Here, the probability that the initial
leader retains the lead decays very slowly in time with an unusual
decay law.

To understand the fate of the initial leader, we need to understand
the degree distribution of the first node.  We can straightforwardly
determine this degree distribution analytically for the constant and
linear attachment rates \cite{kr,DMS2}.  Let $P(k,N)$ be the
probability that the first node has degree $k$ in a network of $N$
links\footnote{The normalized attachment probability is $A_k/A$, with
$A=\sum A_j N_j$.  For the linear attachment rate, $A$ is twice the
total number of links.  Hence formulae are neater if we denote by $N$
the total number of links.}.  For $A_k=k$, this probability obeys
\cite{kr}
\begin{equation}
\label{P}
P(k,N+1)= {k-1\over 2N}\, P(k-1,N)+{2N-k\over 2N}\,\,P(k,N).
\end{equation}
The first term on the right accounts for the case that the earliest
node has degree $k-1$.  Then a new node attaches to it with
probability $(k-1)/2N$, thereby increasing the probability for the
node to have degree $k$.  Conversely, with probability $(2N-k)/2N$ a
new node does not attach to the earliest node, thereby giving the
second contribution to $P(k,N+1)$.

The solution to Eq.~(\ref{P}) for the ``dimer'' initial
condition $\circ\!$\rule[.03in]{.12in}{.01in}$\!\circ$ is
\begin{equation}
\label{Pdimer}
P(k,N)={1\over 2^{2N-k-1}}\, {(2N-k-1)!\over (N-k)!\,(N-1)!}\,.
\end{equation}
For $N\to\infty$, this simplifies to the Gaussian distribution
\begin{equation}
\label{P-final}
P(k,N)\simeq {1\over\sqrt{\pi N}}\,\,e^{-k^2/4N}
\end{equation}
for finite values of the scaling variable $k/N^{1/2}$.  Thus the
typical degree of the first node is of the order of $N^{1/2}$; this is
the same scaling behavior as the degree of the leader node.  For the
trimer initial condition (which we typically used in simulations) we
obtain the degree distribution of the first node as a series of ratios
of gamma functions in which $P(k,N)$ has an $e^{-k^2/4N}$ Gaussian
tail, independent of the initial condition.  The degree of the first
node also approximates that of the leader node more and more closely
as the degree of the first node in the initial state is
increased \cite{MAA02}.

Although $P(k,N)$ contains all information about the degree of the first
node, the behavior of its moments $\langle k^a\rangle_N=\sum k^a P(k,N)$ is
simpler to appreciate.  To determine these moments, it is more convenient to
construct their governing recursion relations directly, rather than to
calculate them from $P(k,N)$.  Using Eq.~(\ref{P}), the average degree of the
first node satisfies the recursion relation $\langle k\rangle_{N+1}=\langle
k\rangle_{N}\left(1+{1\over 2N}\right)$ whose solution is
\begin{equation}
\label{k-soln} \langle
k\rangle_N=\Lambda\,{\Gamma\left(N+\frac{1}{2}\right)\over
{\Gamma\left(\frac{1}{2}\right)\Gamma(N)}}\simeq
{\Lambda\over\sqrt{\pi}}\, N^{1/2}\,.
\end{equation}
The prefactor $\Lambda$ depends on the initial conditions, with $\Lambda=2,
8/3, 16/5,\ldots$ for the dimer, trimer, tetramer, {\it etc.}, initial
conditions.

This multiplicative dependence on the initial conditions means that
the first few growth steps substantially affect the average degree of
the first node.  For example, for the dimer initial condition, the
average degree of the first node is, asymptotically, $\langle
k\rangle_N\simeq 2\sqrt{N/\pi}$.  However, if the second link attaches
to the first node, an effective trimer initial condition arises and
$\langle k\rangle_N\simeq(8/3)\sqrt{N/\pi}$. Thus, small initial
perturbations lead to huge differences in the degree of the first
node.

\begin{figure}[ht]
\centerline{\includegraphics[width=6.5cm]{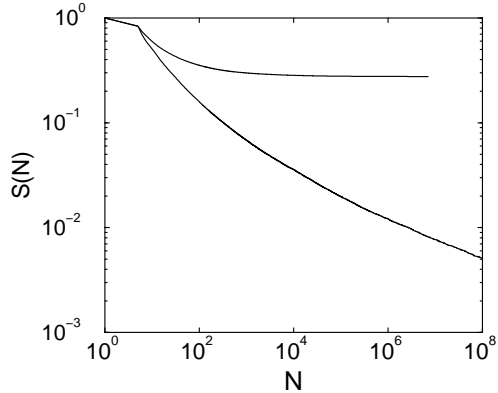}}
\caption{The probability that the first node leads throughout the
evolution obtained from $10^5$ realizations of up to size $N=10^7$ for
$A_k=k$ (upper), and up to $N=10^8$ for $A_k=1$ (lower).}
\label{surv}
\end{figure}

An intriguing manifestation of the rich get richer phenomenon is the
behavior of the survival probability $S(N)$ that the first node leads
throughout the growth up to size $N$ (Fig.~\ref{surv}).  For the
linear attachment rate, $S(N)$ saturates to a finite non-zero value of
approximately 0.277 as $N\to\infty$; saturation also occurs for the
general attachment rate $A_k=k+\lambda$.  We conclude that for
popularity-driven systems, the rich get richer holds in a strong
form---the lead never changes with a positive probability.

For constant attachment rate, $S(N)$ decays to zero as $N\to\infty$,
but the asymptotic behavior is not apparent even when $N=10^8$.  A
power law $S(N)\sim N^{-\phi}$ is a reasonable fit, but the local
exponent is still slowly decreasing at $N\approx 10^8$ where it has
reached $\phi(N)\approx 0.18$.  To understand the slow approach to
asymptotic behavior, we study the degree distribution of the first
node.  This quantity satisfies the recursion relation
\begin{equation}
\label{P0}
P(k,N)={1\over N}\,P(k-1,N-1)+{N-1\over N}\,P(k,N-1)
\end{equation}
which reduces to the convection-diffusion equation
\begin{equation}
\label{Pcont}
\left({\partial \over\partial \ln N}+{\partial \over\partial k}\right)
P={1\over 2}\,{\partial^2 P\over\partial k^2}
\end{equation}
in the continuum limit.  The solution is a Gaussian
\begin{equation}
\label{Pgauss}
P(k,N)\simeq{1\over \sqrt{2\pi \ln N}}\,
\exp\left[-{(k-\ln N)^2\over 2\ln N}\right].
\end{equation}
Therefore the degree of the first node grows as $\ln N$, with
fluctuations of the order of $\sqrt{\ln N}$.  On the other hand, the
maximal degree grows faster, as $\ln N/\ln 2$, with negligible
fluctuations.

We now estimate the large-$N$ behavior of $S(N)$ as $\sum_{k\geq
    k_{\rm max}}P(k,N)$.  This approximation gives
\begin{eqnarray}
\label{phi}
S(N)\sim  \int_{v\ln N}^\infty {dk\over \sqrt{\ln N}}
\exp\left[-{(k-\ln N)^2\over 2\ln N}\right]
\sim N^{-\phi} \,\,(\ln N)^{-1/2}\, ,
\end{eqnarray}
with $\phi=[(\ln 2)^{-1}-1]^2/2\cong0.097989$.  The logarithmic factor
leads to a very slow approach to asymptotic behavior.

The above estimate is based on a Gaussian approximate for $P(k,N)$
which is not accurate for $|k-\ln N|\gg \sqrt{\ln N}$.  However, we
can determine $P(k,N)$ exactly because its defining recursion formula,
Eq.~(\ref{P0}), is closely related to the Stirling numbers ${N\brack
k}$ of the first kind \cite{stir}. For the dimer initial condition,
the solution reads $P(k,N)={N\brack k}/N!$. The corresponding
generating function is \cite{stir}
\begin{eqnarray}
S_N(x)=\sum_{k=1}^N P(k,N)\,x^k={x(x+1)\ldots(x+N-1)\over N!}\,.
\end{eqnarray}
Using the Cauchy theorem, we express $P(k,N)$ in terms of the contour
integral $S_N(x)/x^{k+1}$.  When $N\to\infty$, this contour integral
is easily computed using the saddle point technique.  Finally, we
arrive at Eq.~(\ref{phi}) with the same logarithmic prefactor but with
the slightly smaller {\em exact} transcendental exponent
$\phi=1-\frac{1+\ln\ln 2}{\ln 2}\cong0.08607$.  The remarkably small
exponent value and the logarithmic correction are the reasons why
simulations with $N=10^8$ observed an exponent that was more that
twice larger.

\section{Summary and Discussion}

Extremal properties provide an important statistical characterization
of random structures and these properties yield many insights and
surprises.  Generally, extremes involve logarithmic dependences on
system size.  The practical consequences are numerous: slow
convergence to asymptotic behavior, significant statistical
fluctuations, erratic changes in extremal characteristics, and
sensitive dependence on the initial conditions.  Such behavior is
consistent with our experience.  For example, changes in athletic
records are rare and unpredictable.  As another example, the number of
changes in the composition of the bellwether Dow Jones stock index
(the 30 largest companies) ranged from a high of 11 in the 1990's to a
low of 0 in the 1950's \cite{djia}.

Leadership statistics of random graphs and random trees are quite
similar: lead changes are infrequent; their total number increases
logarithmically with the system size.  The time-dependent number of
lead changes approaches a self-similar form. The convergence to the
asymptotic behavior is much slower for extremal statistics compared
with size statistics because of the presence of various logarithmic
dependences.  Hence, the asymptotic behavior is difficult to detect in
practice, especially for random graphs.

The most elementary leadership characteristic is the overall number of
lead changes as a function of system size.  This quantity can be
measured simply by counting the number of changes until the process
ends, making no reference to time.  We have seen that introducing the
time variable and treating the merger process dynamically not only
produces this quantity, but also reveals an important self-similar
behavior throughout the growth process.

Lead changes are also rare in popularity-driven network growth
processes, where leadership is restricted to the earliest nodes.  With
finite probability, the first node remains the leader throughout the
evolution.  For growth with no popularity bias, leadership is shared
among a somewhat larger cadre of nodes.  As a consequence, the average
index of the leader node grows algebraically with the network size.
The possibility of sharing the lead among a larger subset of nodes
gives a rich dynamics in which the probability that the first node
retains the lead decays algebraically with the system size.

Extremal height properties of random trees can be obtained by
analyzing the underlying nonlinear evolution equations.  The
cumulative distributions of tree heights approach a traveling wave
form and the mean values grow logarithmically with the tree size.  The
corresponding growth coefficients can be obtained using either an
elementary probabilistic argument or using an extremum selection
criteria on the traveling wave. The same formalism used to analyze the
leader and the laggard extends naturally to extremal statistics of
extremal characteristics such as the heights of the tallest and the
shortest trees.

To obtain leader or laggard characteristics, we employed the scaling
behavior of the size distribution outside the scaling regime, namely,
at sizes much larger than the typical size where, at least formally,
statistical fluctuations can no longer be ignored.  Nevertheless, the
size dependences for these various leadership statistics appear to be
asymptotically exact.  Further analysis is needed to illuminate the
role of statistical fluctuations, for example, by characterizing
corrections to the leading behavior \cite{aal,jls,ve}.

\section{Acknowledgments}
We are thankful to our collaborator Satya Majumdar.  This research
was supported by DOE(W-7405-ENG-36) and NSF(DMR0227670).

\end{document}